\documentclass[prd,superscriptaddress,amsfonts,amssymb,amsmath,showpacs,twocolumn,nofootinbib]{revtex4-2}
\usepackage{amsthm}
\usepackage{bm}
\usepackage{amsfonts}
\usepackage{latexsym}
\usepackage{graphicx}
\usepackage{amsmath}
\usepackage{palatino}
\usepackage{xcolor} 
\usepackage{mathpazo}
\usepackage{xcolor}
\usepackage{rotating}
\usepackage{adjustbox}
\usepackage{tensor}
\usepackage{textcomp}
\usepackage{float}
\usepackage{booktabs}
\usepackage{dcolumn}
\usepackage[titletoc]{appendix}
\usepackage{booktabs}
\usepackage{multirow}
\usepackage{hyperref}
\hypersetup{colorlinks,citecolor=blue}
\usepackage{amsmath}
\usepackage{xcolor}
\usepackage{orcidlink}
\usepackage{epsfig}
\usepackage{caption}
\usepackage{subcaption}\newtheorem{theorem}{Theorem}
\usepackage{commath}
\hypersetup{colorlinks,citecolor=blue}
\hypersetup{colorlinks=true,linkcolor=magenta,filecolor=magenta,    urlcolor=blue}

\def\be{\begin{equation}}
\def\ee{\end{equation}}
\def\bea{\begin{eqnarray}}
\def\eea{\end{eqnarray}}

\definecolor{owngreen}{rgb}{0.0, 0.5, 0.0}

\begin{document}

\title{Optical Characteristics and Frequency Shifts of Quartic Wormholes in $f(T,\tau)$ Gravity}

\author{Sourav Chaudhary\,\orcidlink{0009-0008-9590-4761}}
\email{chaudharysourav192@gmail.com}
\affiliation{ Department of Mathematics, 
School of Chemical Engineering and Physical Sciences
Lovely Professional University
Jalandhar, Punjab - 144411, India}

\author{S. K. Maurya \orcidlink{0000-0003-4089-3651}}
\email{sunil@unizwa.edu.om}
\affiliation{Department of Mathematics and Physical Science,
College of Arts and Science, University of Nizwa, Nizwa 616, Sultanate of Oman}
\author{Jitendra Kumar\,\orcidlink{0000-0002-8473-433X}}
 \email{jitendark@gmail.com}
\affiliation{Department of Mathematics, Central University of Haryana, Jant-Pali, Mahendergarh}
\begin{abstract}
This work explores traversable wormhole solutions characterized by a quartic inverse-power density profile within \(f(T,\tau)\) gravity. A specific inverse-power density profile, characterized by proportional to \(1/r^4\)    is introduced to establish the corresponding exact form of the shape function. The derived configuration exhibits all the necessary geometrical features of a traversable wormhole, such as throat formation, flaring-out behavior, and acceptable asymptotic characteristics. To gain further insight into the matter sector, we analyze the corresponding radial profiles of the energy density and pressure components. The throat region is characterised by negative radial pressure and null energy condition violation, both of which are relevant to the exotic matter content of the wormhole solution. To examine the lensing characteristics of the obtained spacetime, we employ the Gauss–Bonnet formalism to evaluate the gravitational deflection angle. Our findings indicate that the parameters \(\gamma\) and \(\rho_0\) exert a considerable influence on the resulting light trajectories. The perturbative behavior of the obtained configuration is investigated within the scalar and electromagnetic field sectors. Subsequently, a novel electromagnetic potential is introduced to examine its influence on the radial variation of the red-blue shifts characteristics. Our findings establish a unified picture of the geometrical, perturbative, and optical characteristics of quartic density-supported wormholes.
\end{abstract}

\maketitle

\section{Introduction}\label{sec1}
Wormholes are theoretical structures in spacetime that could connect two separate regions of the Universe through a geometric shortcut. Although their existence has not yet been confirmed through observations, the physical mechanisms responsible for their formation and the conditions required for their stability remain active areas of research. In particular, the prospect of realizing traversable wormholes has generated significant interest, including the possibility of reproducing wormhole-like systems in laboratory environments in the future. Developments in quantum mechanics added a new dimension to this subject by establishing possible connections between traversable wormholes and quantum information processes. Consequently, several investigations have explored the possibility of experimentally realizing wormhole analogues and utilizing them in quantum teleportation protocols~\cite{Brown:2019hmk,Nezami:2021yaq}. Recent developments have further highlighted the significance of light-ring (LR) structures in generic wormhole geometries, offering a geometrical perspective on their gravitational and optical characteristics~\cite{Xavier:2024iwr}. Recent studies have examined wormhole configurations in galactic halo settings, highlighting possible relationships between their geometrical structures and dark matter distributions~\cite{Mustafa:2023kqt,Mustafa:2022fxn}. The role of solitonic quantum wave dark matter in shaping thin-shell wormhole configurations has also been examined~\cite{Mustafa:2023qxf}. In addition to conventional wormhole models, investigations have also addressed worldsheet traversable wormholes~\cite{Jahnke:2019gxr} and the possibility of configurations that could, in principle, permit human passage~\cite{Maldacena:2020sxe}. The observational aspects of wormhole geometries have emerged as an important area of investigation, particularly concerning their shadow characteristics and gravitational lensing phenomena, which can be examined within the framework of the Karmarkar condition~\cite{Molla:2024loc}. Wormhole solutions have also been explored in braneworld scenarios, where they can potentially establish connections between distinct branes through a higher-dimensional bulk, providing perspectives on quantum-gravity phenomena~\cite{Bronnikov:2002rn,Wang:2017cnd}. The propagation of light in wormhole spacetimes and the associated deflection angles have been examined in~\cite{Aounallah:2020rlf,Zhang:2024sgs}, while their shadow structures have been investigated in~\cite{Huang:2023yqd}. The gravitational lensing properties of wormholes and black holes in a Kasner background were comparatively investigated by Muniz et al.~\cite{Muniz:2026qmb}.  Other important characteristics of wormhole spacetimes have also been discussed in earlier investigations~\cite{Jaramillo:2023pny,Nojiri:2023dvf}.\\

A complementary approach to the geometric description of gravity is provided by teleparallelism, in which torsion rather than curvature plays the central geometrical role. The corresponding torsion scalar \(T\) has therefore become an important quantity in modified theories of gravity~\cite{Capozziello:2011hj,Krssak:2018ywd}. The resulting \(f(T)\) framework, widely known as teleparallel gravity, can be traced back to Einstein's efforts toward a unified theory of gravity and electromagnetism. Unlike the conventional curvature-based formulation, teleparallel gravity employs a curvature-free linear connection defined through the tetrad field. This formalism has been successfully applied to the construction and analysis of different wormhole solutions~\cite{Boehmer:2012uyw,Jamil:2011ptc}. Subsequently, Harko et al.~\cite{Harko:2014aja} proposed an extended teleparallel framework in which the matter sector is explicitly coupled to the torsional geometry, leading to the \(f(T,\tau)\) formalism. The viability of teleparallel gravity has further been examined against observational datasets, including SNIa, BAO, and CMB observations~\cite{Wu:2010mn}. Moreover, constraints derived from cosmic chronometers and  observations have provided additional support for investigating these theories~\cite{Nunes:2018evm}. The versatility of teleparallel gravity has led to its application across a broad range of cosmological and astrophysical problems. In particular, teleparallel dark-energy models have been employed to describe the late-time accelerated expansion of the Universe, with solutions exhibiting both quintessence-like and phantom-like behavior~\cite{Geng:2011aj,Wei:2011yr}. Such models have consequently been considered as possible alternatives to the conventional CDM framework. Beyond cosmological applications, the behavior of gravitational waves in teleparallel theories has also been investigated in the weak-field limit~\cite{Bamba:2013ooa}. An important extension arises when the torsional geometry is coupled directly to the matter sector. Within the \(f(T,\tau)\) framework, the squared-torsion model has been confronted with observational constraints~\cite{Arora:2022jso}. Rebou\c{c}as et al.~\cite{Reboucas:2026fxc} studied the optical behaviour of reflection-asymmetric polymer thin-shell wormholes surrounded by one-sided Kiselev-type quintessence. Regarding wormhole geometries, Ditta et al.~\cite{Ditta:2021uoe} investigated solutions satisfying the Karmarkar condition in both linear and nonlinear \(f(T,\tau)\) gravity. In a related study, Mustafa et al.~\cite{Mustafa:2021vqz} examined wormhole solutions sourced by Gaussian and Lorentzian density distributions within a linear \(f(T,\tau)\) model. In a recent study, Muniz et al.~\cite{Muniz:2026akg} investigated hyperbolic wormhole geometries sustained by a Casimir-memory source and analysed their shell-free matching conditions. Rebou\c{c}as  et al.~\cite{Reboucas:2026fwz} presented a comprehensive investigation of polymer-quintessence thin-shell wormholes, addressing their stability, thermodynamic behaviour, optical geometry, and observable features. These investigations demonstrate the growing relevance of matter-coupled teleparallel gravity in describing nontrivial wormhole geometries and motivate further studies of their gravitational and optical properties.

\section{ $f(T,\tau)$ Gravity  }\label{sec2}
The geometric framework of teleparallel gravity is described in terms of the tetrad field ($e^{\mathcal{A}}_{\ i}$), which serves as the fundamental field from which the spacetime metric is defined as
\begin{equation}
g_{ij}=\eta_{\mathcal{AB}}e^{\mathcal{A}}_{\ i}e^{\mathcal{B}}_{\ j},
\qquad
e\equiv\det(e^{\mathcal{A}}_{\ i})=\sqrt{-g},
\end{equation}
with
\begin{equation}
\eta_{\mathcal{AB}}=
\begin{pmatrix}
-1 & 0 & 0 & 0\\
0 & 1 & 0 & 0\\
0 & 0 & 1 & 0\\
0 & 0 & 0 & 1
\end{pmatrix}.
\end{equation}
Within the pure-tetrad formulation, the spin connection is assumed to vanish, i.e., $\omega^{\mathcal{A}}_{\mathcal{ B}i}=0$, and the torsion tensor consequently takes the form
\begin{equation}
T^{\lambda}_{\ ij}
=
e_{\mathcal{A}}^{\ \lambda}
\left(
\partial_{i}e^{\mathcal{A}}_{\ j}
-
\partial_{j}e^{\mathcal{A}}_{\ i}
\right).
\end{equation}

The geometrical quantities associated with the torsional description, namely the contorsion tensor and the superpotential, are defined respectively by
\begin{equation}
K^{ij}_{\ \ \rho}
=
-\frac{1}{2}
\left(
T^{ij}_{\ \ \ \rho}
-
T^{ji}_{\ \ \ \rho}
-
T_{\rho}^{\ \ ij}
\right),
\end{equation}
and
\begin{equation}
S_{\rho}^{\ ij}
=
\frac{1}{2}
\left(
K^{ij}_{\ \ \rho}
+
\delta_{\rho}^{i}T^{\alpha j}_{\ \ \alpha}
-
\delta_{\rho}^{j}T^{\alpha i}_{\ \ \alpha}
\right),
\end{equation}
so that the torsion scalar is
\begin{equation}
T=S_{\rho}^{\ ij}T^{\rho}_{\ ij}.
\end{equation}

Within the \(f(T,\tau)\) framework, the gravitational action is defined by
\begin{equation}\label{e7}
S=
\frac{1}{16\pi}
\int d^{4}x\,e\,f(T,\tau)
+
\int d^{4}x\,e\,\mathcal{L}_{m}.
\end{equation}

For the anisotropic matter configuration considered in the present study
\begin{equation}
\varsigma^{i}_{\ j}
=
\mathrm{diag}(-\rho,\mathcal{P}_{r},\mathcal{P}_{t},\mathcal{P}_{t}).
\end{equation}

The matter Lagrangian is chosen to correspond to the mean pressure
\begin{equation}
\mathcal{L}_{m}
=
\mathcal{P}
=
\frac{\mathcal{P}_{r}+2\mathcal{P}_{t}}{3},
\end{equation}
which naturally reduces to $\mathcal{L}_{m}=\mathcal{P}$ for an isotropic matter distribution. Consistent with the conventional perfect-fluid description of trace-coupled gravity theories, we employ the following closure prescription
\begin{equation}
\frac{\partial^{2}\mathcal{P}}
{\partial g^{ij}\partial g^{\alpha\beta}}
=0.
\end{equation}
The corresponding matter variation takes the form
\begin{equation}
\varsigma_{ij}^{(\mathrm{var})}
=
-2\varsigma_{ij}
+
\mathcal{P}g_{ij}.
\end{equation}

Taking the variation of Eq.~(\ref{e7}) with respect to the tetrad, we obtain
\begin{eqnarray}\label{e12}
 &&\hspace{-0.5cm}   e_{\mathcal{A}}^{\rho}S_{\rho}^{\ ij}
(\partial_{i}T)f_{TT}
+
e_{\mathcal{A}}^{\rho}S_{\rho}^{\ ij}
(\partial_{i}\tau)f_{T\tau}
+
e^{-1}\partial_{i}
\left(
e e_{\mathcal{A}}^{\rho}S_{\rho}^{\ ij}
\right)f_{T}
\nonumber\\ && \hspace{-0.5cm}+
e_{\mathcal{A}}^{i}T^{\lambda}_{\ i\kappa}
S_{\lambda}^{\ j\kappa}f_{T}
-
\frac{1}{4}e_{\mathcal{A}}^{j}f
-
\frac{f_{T}}{2}
\left(
e_{\mathcal{A}}^{\lambda}\varsigma_{\lambda}^{\ j}
+
\mathcal{P}e_{\mathcal{A}}^{j}
\right)
\nonumber\\ && \hspace{-0.5cm}=
-4\pi e_{\mathcal{A}}^{\lambda}\varsigma_{\lambda}^{\ j}.
\end{eqnarray}

With the linear functional form \(f(T,\tau)=T+\gamma\,\tau\), the field equations can be written as
\begin{equation}\label{e14}
G_{ij}
=
(8\pi-\gamma)\Theta_{ij}
-\gamma \mathcal{P}
+\frac{\tau}{2}\delta_{ij},
\end{equation}
where
\begin{equation}
\mathcal{P}+\frac{\tau}{2}
=
\frac{-3\rho+5\mathcal{P}_r+10\mathcal{P}_t}{6}.
\end{equation}

As $\gamma\rightarrow0$, the modified field equations continuously approach the general-relativistic equations. Although the geometric sector remains that of TEGR for the adopted linear form, the trace coupling introduces the modification. A close structural analogy exists with the linear $f(\mathcal{R},T)$ theory~\cite{Harko:2011kv}. Nevertheless, the coefficients obtained from the matter variation in Eq.~(\eqref{e14}) distinguish the present formulation from that framework. When \(\gamma\neq0\), the matter tensor generally does not satisfy the standard conservation law. By taking the covariant divergence of Eq.~(\eqref{e14}) and using the Bianchi identity, we arrive at
\begin{equation}
\nabla^i\Theta_{ij}
=
\frac{\gamma}{8\pi-\gamma}
\nabla_j\left(\mathcal{P}+\frac{\tau}{2}\right).
\end{equation}

The chosen linear model admits a consistent diagonal tetrad. Under static spherical symmetry, the off-diagonal equation reduces to
\begin{equation}
f_{TT}T' + f_{T\tau}\tau' = 0,
\end{equation}
which is identically satisfied under the conditions $f_{TT}=f_{T\tau}=0$.
It follows that the diagonal tetrad in spherical coordinates is compatible
with the pure-tetrad description of the present linear model. Even in flat spacetime, the adopted tetrad may exhibit a nonvanishing
inertial torsion contribution. However, the linear form of the model,
characterized by $f_{TT}=f_{T\tau}=0$, ensures that this contribution
imposes no additional field-equation constraint.\\

The wormhole geometry is characterized by the Morris-Thorne line element.
\begin{equation}
ds^2
=
-e^{2\Phi(r)}dt^2
+\frac{dr^2}{1-{b(r)/r}}
+r^2d\theta^2
+r^2\sin^2\theta\,d\phi^2,
\label{e18}
\end{equation}
where $\Phi(r)$ and $b(r)$ respectively govern the the gravitational redshift and shape of the wormhole.\\

The wormhole throat is located at the minimum radial coordinate ($r=r_0$), where the shape function obeys
\begin{equation}
b(r_0)=r_0.
\end{equation}
For the geometry to satisfy the flaring-out requirement, the derivative of the shape function at the throat must fulfil
\begin{equation}
b'(r_0)<1,
\qquad
b(r)<r,\quad r>r_0.
\label{eq:18}
\end{equation}
Furthermore, the redshift function $\Phi(r)$ should remain finite throughout the spacetime so that no event horizon develops.\\

At large radial distances, the geometry must approach the Minkowski spacetime. This asymptotic behaviour is ensured by the conditions
\begin{equation}
\lim_{r\rightarrow\infty}\frac{b(r)}{r}=0,
\qquad
\lim_{r\rightarrow\infty}\Phi(r)=\mathrm{constant}.
\end{equation}
The tetrad representation consistent with Eq.~\eqref{e18} can be written as
\begin{equation}
e^{A}{}_{i}
=
\begin{pmatrix}
e^{\Phi} & 0 & 0 & 0\\
0 & \dfrac{1}{\sqrt{1-\dfrac{b}{r}}} & 0 & 0\\
0 & 0 & r & 0\\
0 & 0 & 0 & r\sin\theta
\end{pmatrix}.
\end{equation}
which gives the torsion scalar
\begin{equation}
T(r)
=
-\left(\frac{1}{r}+2\Phi'\right)\times\frac{2}{r}
\left(1-\frac{b}{r}\right)
.
\label{eq:20}
\end{equation}

Using the metric in Eq.~\eqref{e12}, we obtain the corresponding anisotropic source components
\begin{eqnarray}
\label{e23}
&&\hspace{-1cm}\rho = \frac{1}{\mathcal{D} r^2}
\Big[
-10\gamma r(r-b)\Phi'^2
+5\gamma\Phi'\left\{3b+r(b'-4)\right\}
\nonumber\\ && \hspace{-0.5cm}-16(\gamma-3\pi)b'
-10\gamma r(r-b)\Phi''
\Big], \\
\label{e24}
&&\hspace{-1cm}\mathcal{P}_r = \frac{1}{\mathcal{D} r^3}
\Big[
10\gamma r^3\Phi'^2
-8\gamma r b'
+r^2\Phi'\left(96\pi-28\gamma-5\gamma b'\right)
\nonumber\\ && \hspace{-0.5cm}+10\gamma r^3\Phi''
-b\Big\{
48\pi-24\gamma
+(96\pi-33\gamma)r\Phi'
\nonumber\\ && \hspace{-0.5cm}+10\gamma r^2\Phi'^2
+10\gamma r^2\Phi''
\Big\}
\Big],\\
\label{e25}
&&\hspace{-1cm}\mathcal{P}_t = \frac{1}{\mathcal{D} r^3}
\Big[
b\Big\{
-12(\gamma-2\pi)
-3(\gamma+8\pi)r\Phi'
\nonumber\\ && \hspace{-0.5cm}+2(7\gamma-24\pi)r^2
\left(\Phi'^2+\Phi''\right)
\Big\}
+r\Big\{
-2(7\gamma-24\pi)r^2\nonumber\\ && \hspace{-0.5cm}\times
\left(\Phi'^2+\Phi''\right)
+4(\gamma-6\pi)b'
\nonumber\\ && \hspace{-0.5cm}+r\Phi'
\left[48\pi-4\gamma+(7\gamma-24\pi)b'\right]
\Big\}
\Big],
\end{eqnarray}
where $\mathcal{D}=24(\gamma-8\pi)(\gamma-2\pi)$.\\

We simplify the geometrical description by adopting the zero-tidal-force approximation, under which the redshift function remains constant. Without loss of generality, its constant value can be absorbed into the time coordinate, giving $\Phi'=0.$

Therefore Eqs.~(\ref{e23}-\ref{e25}) becomes
\begin{eqnarray}
&&\hspace{-0.5cm}\rho
=
-\frac{2(\gamma-3\pi)b'}
{3(\gamma-8\pi)(\gamma-2\pi)r^2},\\
&&\hspace{-0.5cm}\mathcal{P}_r
=
\frac{3(\gamma-2\pi)b-\gamma r b'}
{3(\gamma-8\pi)(\gamma-2\pi)r^3},\\
&&\hspace{-0.5cm}\mathcal{P}_t
=
\frac{-3(\gamma-2\pi)b+(\gamma-6\pi)r b'}
{6(\gamma-8\pi)(\gamma-2\pi)r^3}.
\end{eqnarray}
  \begin{figure*}[!htp]
    \centering
\includegraphics[height=5.3cm,width=5cm]{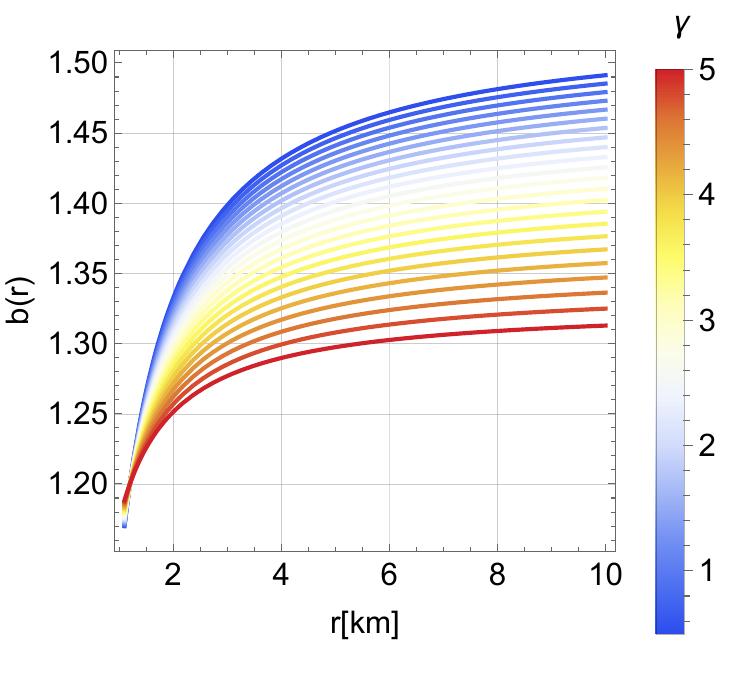}~~~~ \includegraphics[height=5.3cm,width=5cm]{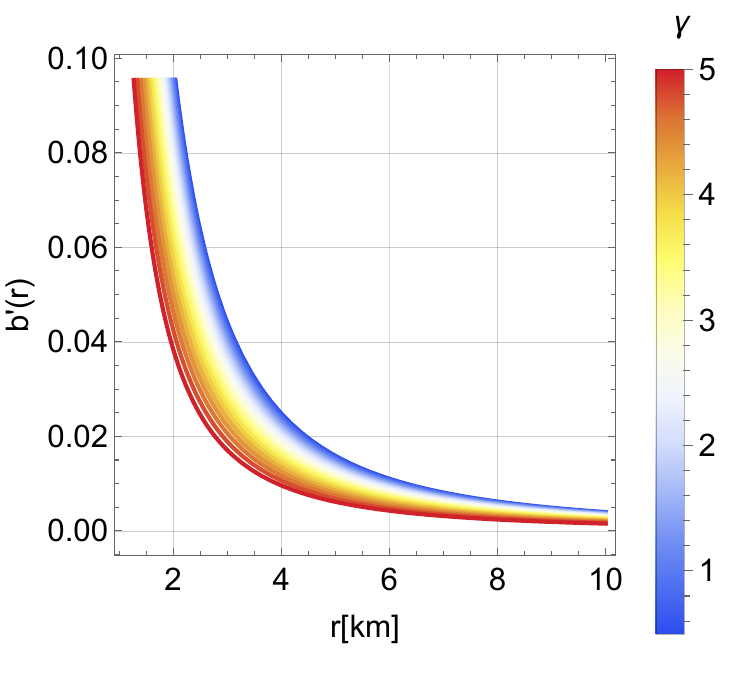}~~~~\includegraphics[height=5.3cm,width=5cm]{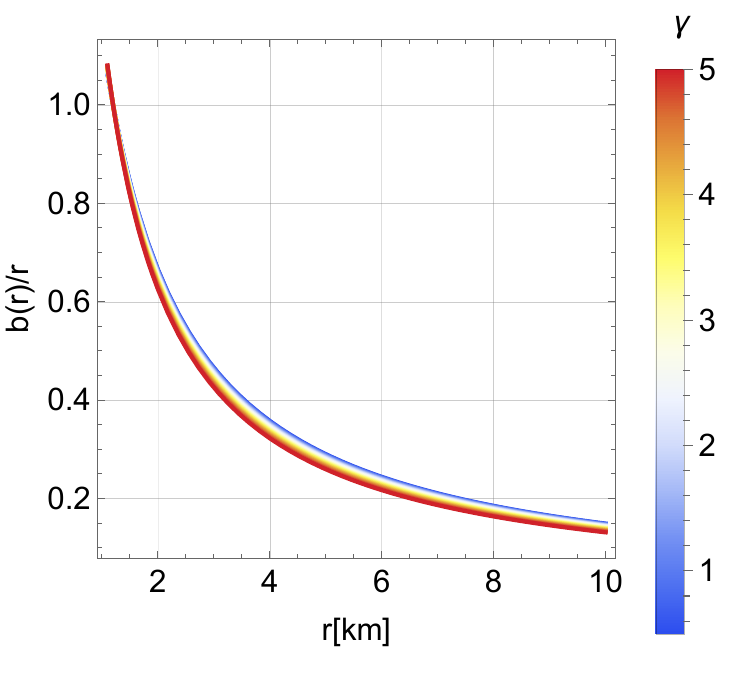}
    \caption{Shape function and its geometrical properties for $\rho_0=0.008$ and $r_0=1.2$: (left)  $b(r)$, (center) flaring-out behavior at $r=r_0$, and (right) asymptotically flat as $r\rightarrow\infty$.}
    \label{fig2a}
\end{figure*}

\section{ Wormhole Solutions with $f(T,\tau)=\gamma~\tau+T$}\label{sec6}
The quartic inverse-power density profile is a phenomenologically motivated model for describing radially decreasing matter distributions in spherically symmetric gravitational systems. Its mathematical structure is characterized by an inverse fourth-power dependence on the radial coordinate, expressed as $\rho(r)\propto r^{-4}$, indicating that the density decreases rapidly as the distance from the central region increases. More specifically, the profile can be written as~\cite{William}
\begin{eqnarray}
&&\hspace{-0.cm} \rho=\rho_0\Big(r_0/r\Big)^4,
\end{eqnarray}
where $\rho_0$ represents the density at a throat radius $r_0$. Inverse-power laws are widely adopted in mathematical physics and astrophysics for their ability to describe spatially varying physical quantities in a simple analytical form. In spherical systems, such profiles are particularly relevant because the radial density distribution determines the enclosed matter content through the integral relation $M(r)=4\pi~\int_{r_{0}}^{r}\rho(\bar{r})\bar{r^2}d\bar{r}$. For the quartic inverse-power profile, the density scales as $r^{-4}$, whereas the mass contribution per unit radial interval scales as $dM/dr\propto r^{-2}$. This ensures that the integrated matter contribution over the exterior spatial domain is finite, since the corresponding radial integral,  $\int_{r_0}^{\infty}r^{-2},dr$ converges. These considerations establish the relevance of this profile to traversable wormhole configurations, for which a consistent analysis requires examining the matter sector together with the geometrical constraints at the throat and the behaviour of spacetime at large distances. The incorporation of a rapidly decaying quartic inverse-power density profile into the field equations provides a framework for examining its implications for the wormhole geometry, the associated energy-density distribution, and the possibility of achieving asymptotic flatness. Furthermore, the profile provides a useful framework for studying the relationship between matter localization and geometric admissibility without imposing a constant density throughout the spacetime. It should be emphasized, however, that the inverse-fourth-power dependence is adopted here as a modelling ansatz rather than a universally established density law; its physical viability must therefore be evaluated through the resulting field equations, traversable requirements, and relevant energy conditions.\\

On equating wormhole density with quartic density profile, we get differential equation of the form 
\begin{eqnarray}
   -\frac{2(\gamma-3\pi)b'}
{3(\gamma-8\pi)(\gamma-2\pi)r^2} -\rho_0\Big(r_0/r\Big)^4=0.
\end{eqnarray}
On solving the above differential equation for shape function and imposing throat condition, we get
\begin{eqnarray}\label{e31}
 &&\hspace{-0.cm} b(r)=\frac{3 \gamma ^2 r_0^3 \rho_0-30 \pi  \gamma  r_0^3 \rho_0+48 \pi ^2 r_0^3 \rho_0-2 \gamma  r_0+6 \pi  r_0}{2 (3 \pi -\gamma )}\nonumber\\ && \hspace{-0.5cm}-\frac{3 \left(\gamma ^2-10 \pi  \gamma +16 \pi ^2\right) r_0^4 \rho_0}{6 \pi  r-2 \gamma  r}.
\end{eqnarray}

\section{Physical Viability of the Wormhole Configuration}
\subsection{Geometrical Analysis of the Shape Function}
The physical viability of the wormholes can be examined through
the shape function $b(r)$, which determines the spatial structure of a
static, spherically symmetric spacetime.
A physically admissible wormhole must satisfy several fundamental
geometric conditions, including the throat condition $b(r_0)=r_0$,
the flaring-out condition $b'(r_0)<1$, and the exterior condition
$b(r)<r$ for $r>r_0$. Furthermore, asymptotic flatness requires
\begin{equation}
\lim_{r\rightarrow\infty}\frac{b(r)}{r}=0.
\end{equation}
By differentiating Eq.~(\ref{e31}) with respect to $r$, we find
\begin{equation}
b'(r)=
\frac{
3r_0^4\rho_0
(\gamma^2-10\pi\gamma+16\pi^2)
}{
2(3\pi-\gamma)r^2
}.
\end{equation}
Thus, at the throat, the flaring-out condition is expressed as
\begin{equation}
b'(r_0)=
\frac{
3r_0^2\rho_0
(\gamma^2-10\pi\gamma+16\pi^2)
}{
2(3\pi-\gamma)
}<1.
\end{equation}
For the fixed parameter values $\rho_0=0.008$, $r_0=1.2$, and
$\beta=0.8$, with $0.5\leq\gamma\leq5$, the derivative at the throat
remains below unity, satisfying the local flaring-out requirement
throughout the specified interval. It is worth noting that the parameter
$\gamma$ explicitly appear in the shape function and
therefore directly influence these geometric conditions.
To investigate asymptotic flatness, we express the shape function as
\begin{equation}
b(r)=A(\gamma)-\frac{C(\gamma)}{r},
\end{equation}
where
\begin{equation}
A(\gamma)=
\frac{
3\gamma^2r_0^3\rho_0
-30\pi\gamma r_0^3\rho_0
+48\pi^2r_0^3\rho_0
-2\gamma r_0+6\pi r_0
}{
2(3\pi-\gamma)
}
\end{equation}
and
\begin{equation}
C(\gamma)=
\frac{
3r_0^4\rho_0
(\gamma^2-10\pi\gamma+16\pi^2)
}{
2(3\pi-\gamma)
}.
\end{equation}
Since $A(\gamma)$ and $C(\gamma)$ are independent of $r$, we have
\begin{equation}
\frac{b(r)}{r}
=\frac{A(\gamma)}{r}
-\frac{C(\gamma)}{r^2},
\end{equation}
which gives
\begin{equation}
\lim_{r\rightarrow\infty}\frac{b(r)}{r}=0.
\end{equation}
Consequently, the radial behaviour of the shape function is consistent with an asymptotically flat geometry.\\

Moreover,
using the throat identity $b(r_0)=r_0$, the exterior difference can be
shown as
\begin{equation}
r-b(r)=
\frac{(r-r_0)}{r}\left(r-\dfrac{C(\gamma)}{r_0}\right).
\end{equation}
Therefore, if $C(\gamma)/r_0<r_0$, then
\begin{equation}
r-b(r)>0,\qquad r>r_0,
\end{equation}
which ensures that
\begin{equation}
b(r)<r,\qquad
1-\frac{b(r)}{r}>0,\qquad r>r_0.
\end{equation}
These inequalities guarantee the required exterior geometry without
an additional zero of $1-b(r)/r$ in the exterior region. When the throat condition is consistently maintained by the chosen parameters, the resulting spacetime exhibits the essential geometrical features required for traversable wormhole solutions within the considered range. A thorough physical analysis also requires checking the matter field equations, the relevant energy conditions, and other physical constraints.

\subsection{Analysis of Energy Conditions }
 \begin{figure*}[!htp]
    \centering
\includegraphics[height=8.3cm,width=8.7cm]{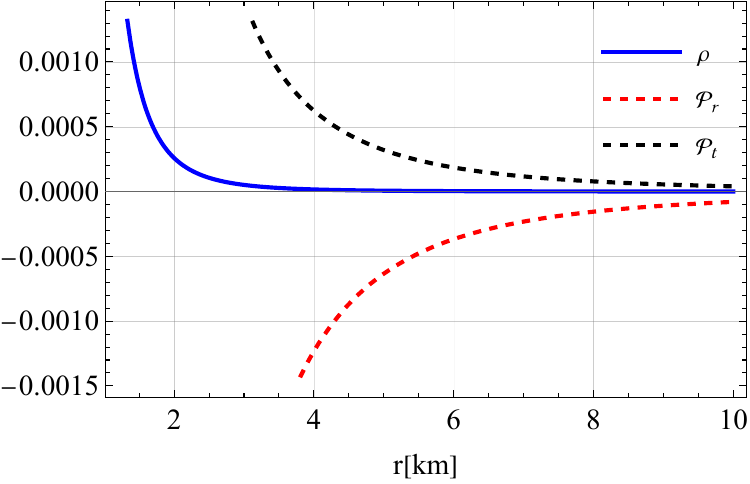}
    \caption{Variation of energy density, radial pressure, and tangential pressure with radial distance for $\rho_0=0.008$, $\gamma=1.2$ and $r_0=1.2$}
    \label{fig1a}
\end{figure*}
Keeping the parameters $\rho_0=0.008$ and $r_0=1.2$ fixed, we consider different values of $\gamma$ in the interval $0.5\leq\gamma\leq5$ and examine the resulting changes in the radial profiles of energy density and pressure. The energy density and pressure components exhibit different radial dependencies, while the distinction between radial and tangential pressures reflects the anisotropic nature of the matter source. The negative radial pressure plays an essential role in maintaining the throat against gravitational contraction, whereas the difference between the radial and tangential pressure components determines the extent of matter anisotropy. The changes in the energy-density and pressure profiles across different values of \(\gamma\) demonstrate how this parameter affects the matter distribution and the associated wormhole geometry. The above analysis provides the necessary background for evaluating the energy conditions and determining whether the resulting wormhole configuration is physically admissible.\\

At the wormhole throat \(r=r_0\), the radial and tangential null energy conditions (NECs) are investigated by substituting the parameter values \(\rho_0=0.008\), \(r_0=1.2\), and \(0.8\leq\gamma\leq5\). For the chosen parameters, we obtain
\[
r_0^2\rho_0=(1.2)^2(0.008)=0.01152.
\]

Substituting this value into the radial null energy expression yields
\[
\begin{aligned}
(\rho+\mathcal{P}_r)|_{r={r_0}}
&=\frac{\mathcal{N}_r(\gamma)}
{2.88(2\pi-\gamma)(3\pi-\gamma)(8\pi-\gamma)},
\end{aligned}
\]
where
\[
\begin{aligned}
\mathcal{N}_r(\gamma)={}&-\gamma^2(0.03456\gamma+2)
+12\pi\gamma(0.03456\gamma+1)\\
&-18\pi^2(0.06912\gamma+1)
+1.10592\pi^3.
\end{aligned}
\]
Numerical evaluation over the considered parameter interval gives
\[
{(\rho+\mathcal{P}_r)_{r_0}<0,
\qquad 0.8\leq\gamma\leq5.}
\]
Thus, the radial NEC is violated throughout the selected interval.

Similarly, substituting the same parameter values into the tangential null energy expression gives
\[
\begin{aligned}
(\rho+\mathcal{P}_t)_{r_0}
&=\frac{\mathcal{N}_t(\gamma)}
{5.76(2\pi-\gamma)(3\pi-\gamma)(8\pi-\gamma)},
\end{aligned}
\]
where
\[
\begin{aligned}
\mathcal{N}_t(\gamma)={}&3.31776\pi^3
+\pi^2(18-2.9952\gamma)\\
&+\gamma^2(2-0.0576\gamma)\\
&+4\pi\gamma(-3+0.19584\gamma).
\end{aligned}
\]
The numerical results establish that
\[
{(\rho+\mathcal{P}_t)_{r_0}>0,
\qquad 0.8\leq\gamma\leq5.}
\]
Therefore, the tangential NEC remains satisfied throughout the investigated interval.

Furthermore, since
\[
0<\gamma\leq5<2\pi<3\pi<8\pi,
\]
all denominator factors remain positive, and no singularities occur in the considered parameter range. Representative numerical values are listed in Table~\ref{tab:NEC_throat}.\\

\begin{table}[htbp]
\centering
\caption{Null energy conditions in the radial and tangential directions, evaluated at the throat for \(\rho_0=0.008\) and \(r_0=1.2\).}
\label{tab:NEC_throat}
\begin{tabular}{c|cc}
\hline
$\gamma$ & $(\rho+\mathcal{P}_r)_{r_0}$ & $(\rho+\mathcal{P}_t)_{r_0}$\\
\hline
0.8 & $-0.03726$ & $0.03463$\\
2.0 & $-0.04512$ & $0.03856$\\
3.0 & $-0.05527$ & $0.04363$\\
4.0 & $-0.07303$ & $0.05251$\\
5.0 & $-0.11546$ & $0.07373$\\
\hline
\end{tabular}
\end{table}

 \begin{figure*}[!htp]
    \centering
\includegraphics[height=7.3cm,width=7.8cm]{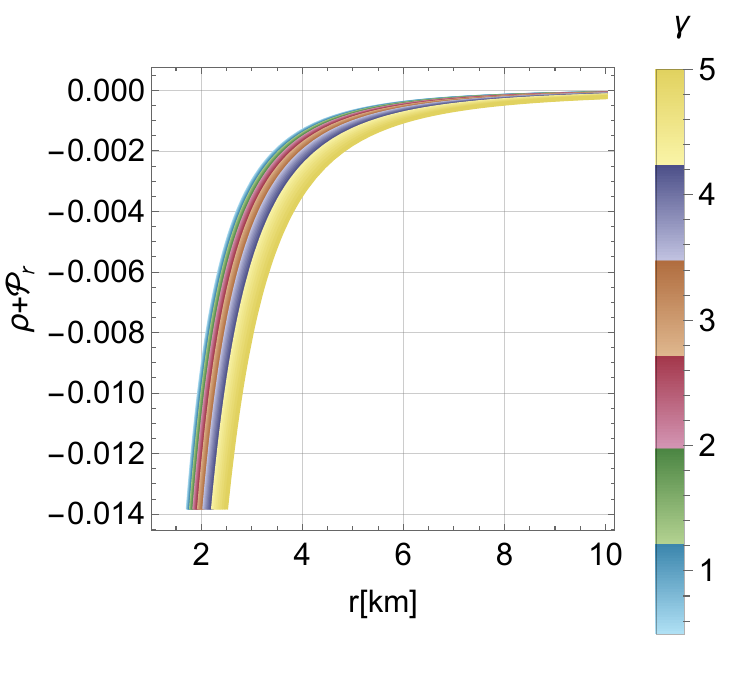}~~~~ \includegraphics[height=7.3cm,width=7.8cm]{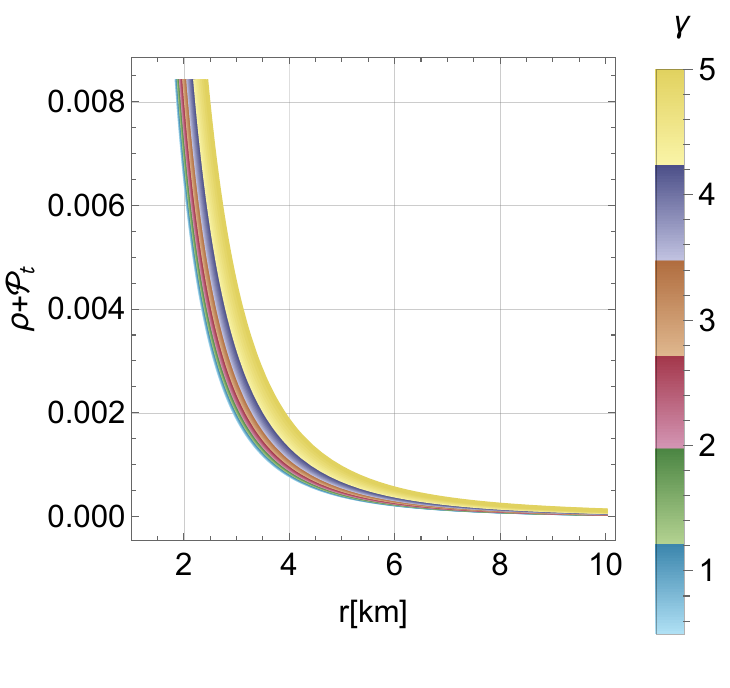}
    \caption{Variation of radial NEC~(left) and tangential NEC~(right) with variation of $\gamma$.}
    \label{fig03}
\end{figure*}
These results indicate that the matter distribution exhibits anisotropic energy-condition behaviour at the throat, with the radial NEC violated and the tangential NEC satisfied. The radial violation in Fig.~\ref{fig03} is consistent with the exotic matter requirement commonly associated with traversable wormhole geometries. The stated sign conditions are based on numerical evaluation over the specified interval.

\section{Deflection Angle }
In this section, we examine the gravitational deflection of light in the spacetime geometry of the obtained wormhole solutions. To evaluate the corresponding deflection angle, we employ the Gauss-Bonnet theorem (GBT), which provides a geometric framework for analyzing the propagation of light in curved spacetime.
\subsection{ Optical Curvature in the Wormhole Geometry}
To investigate the optical properties of the wormhole spacetime, we consider the deflection of light rays along null geodesics. By setting \(ds^2=0\) and considering the equatorial plane \(\theta=\pi/2\), the optical metric corresponding to the Morris-Thorne geometry in Eq.~(\ref{e18}) can be expressed as
\begin{equation}
    dt^2 =
    \frac{1}{1-\dfrac{b(r)}{r}}\,dr^2
    + r^2 d\phi^2.
    \label{eq:46}
\end{equation}
For a zero-tidal-force wormhole, $\Phi(r)=0$, and the temporal
coordinate can be rescaled to absorb the corresponding constant
factor. We next introduce a new radial coordinate $r^*$ to
rewrite the optical metric in terms of the function $f(r^*)$:
\begin{equation}
    dt^2 = \tilde{g}_{ab}dx^a dx^b
    = dr^{*2} + f^2(r^*)d\phi^2,
    \qquad (a,b)=(r^*,\phi).
    \label{eq:47}
\end{equation}
 From Eqs.~\eqref{eq:46} and \eqref{eq:47}, the following relations can be established
\begin{equation}
    dr^* =
    \frac{dr}{\sqrt{1-\dfrac{b(r)}{r}}},
    \qquad
    f(r^*)=r.
    \label{eq:48}
\end{equation}
To examine the curvature properties of the optical geometry, we evaluate the Gaussian curvature \(\mathcal{K}\), expressed as~\cite{Gibbons:2008rj}
\begin{equation}
\mathcal{K}=-\frac{1}{f(r^*)}
\frac{d}{dr^*}
\left[
\frac{df(r^*)}{dr^*}
\right].
\label{eq:49}
\end{equation}
 \begin{figure*}[!htbp]
    \centering
\includegraphics[height=8cm,width=9cm]{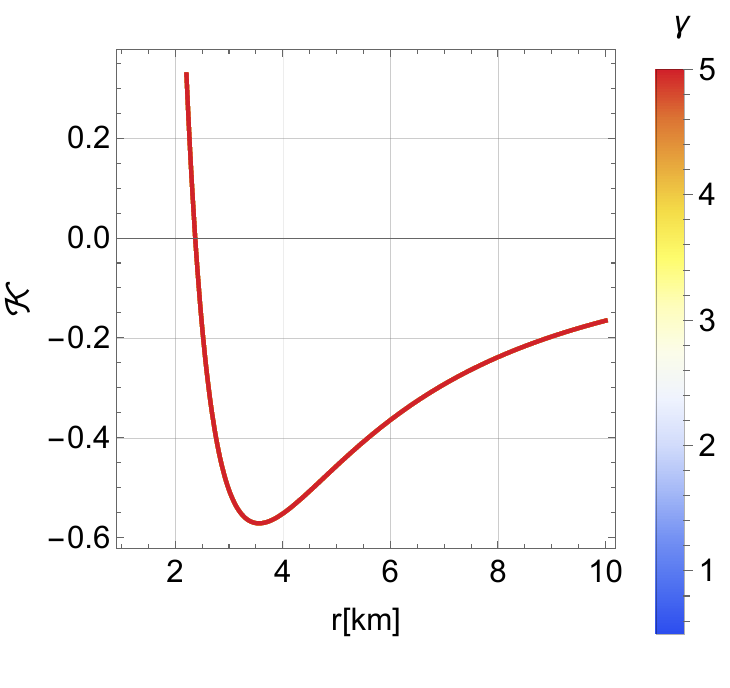}
    \caption{Variation of optical curvature with radial distance for $\rho_0=0.008$, $\gamma=1.2$ and $r_0=1.2$}
    \label{fig4}
\end{figure*}
By employing Eqs.~\eqref{eq:46}-\eqref{eq:47} and carrying out the required simplifications, the Gaussian curvature of the optical metric becomes
\begin{eqnarray}
 &&\hspace{-0.cm}\mathcal{K} = \frac{1}{4 (3 \pi -\gamma ) r^4}\Bigg(-3 \gamma ^2 r_0^3 \rho_0 (r-2 r_0)+2 \gamma  r_0 \Big(15 \pi  r_0^2 \rho_0\nonumber\\ && \hspace{-0.5cm} (r-2 r_0)+r\Big)-6 \pi  r_0 \left(8 \pi  r_0^2 \rho_0 (r-2 r_0)+r\right)\Bigg).
\end{eqnarray}
The resulting optical curvature enables us to explore how the wormhole geometry influences the propagation and deflection of light rays. The Gaussian optical curvature \(\mathcal{K}\) quantifies the intrinsic curvature of the optical geometry associated with the wormhole spacetime. The changes in optical behaviour with increasing \(\gamma\) indicate the role of the coupling parameter in determining the geometrical properties of the wormhole spacetime.\\

The curvature decreases with the radial coordinate according to an
$r^{-4}$ power law, while the parameter $\gamma$ determines its
overall magnitude through the factor $3\pi-\gamma$.  The parameter \(\gamma\) influences the overall curvature, including its possible sign, whereas the radial dependence determines its behaviour as one moves away from the wormhole throat. The plotted profile indicates a rapid decline in \(\mathcal{K}\), followed by a sign change and the appearance of a negative minimum at approximately \(r=3.5\,\mathrm{km}\). The curvature gradually recovers from its minimum and approaches zero at larger distances, although it remains negative throughout the considered interval. Such a variation indicates that the optical curvature changes sign and gradually approaches zero as the radial distance increases. The plotted curves illustrate the effect of \(\gamma\) on the optical geometry. As all considered values satisfy \(\gamma<3\pi\), the denominator does not vanish within the selected parameter range. As shown in Fig.~\ref{fig4}, the parameter \(\gamma\) modifies the optical curvature, which in turn determines an important geometrical aspect of light propagation in the wormhole spacetime.

\subsection{ Gauss-Bonnet Theorem}

\begin{theorem}
Consider a smooth, simply connected domain
$(\mathcal{H}_R,\bar{g})$ in the Riemannian manifold
$(M,\bar{g})$, with its boundary formed by the curve $\eta_R$
and the geodesic segment $\eta_{\bar{g}}$. Let $\mathcal{K}$
and $\kappa$ denote the Gaussian curvature of the manifold
and the geodesic curvature of the boundary
$\partial\mathcal{H}_R=\eta_{\bar{g}}\cup\eta_R$, respectively.
The Gauss-Bonnet theorem (GBT) then reads~\cite{Gibbons:2008rj,Werner:2012rc}:
\begin{equation}
\oint_{\partial\mathcal{H}_R} \kappa\,dt +
\iint_{\mathcal{H}_R} \mathcal{K}\,dS
+
\sum_i \xi_i
=
2\pi\chi(\mathcal{H}_R).
\end{equation}
Here, \(dS\) refers to the differential area element, \(\xi_i\) denotes the exterior angle at each boundary vertex, and \(\chi(\mathcal{H}_R)\) characterises the topology of the domain.
\end{theorem}

For the evaluation of the deflection angle within the
Gauss-Bonnet framework, we make use of the fact that a geodesic
has vanishing geodesic curvature. Accordingly,
$\kappa(\eta_{\bar{g}})=0$ for the geodesic segment
$\eta_{\bar{g}}$. The boundary integral therefore receives
contributions only from the circular arc $\eta_R$, whose
geodesic curvature takes the following form:
\begin{equation}
\kappa(\eta_R)
=
\left|
\nabla_{\dot{\eta}_R}\dot{\eta}_R
\right|.
\end{equation}
  \begin{figure*}
    \centering
\includegraphics[height=7cm,width=8cm]{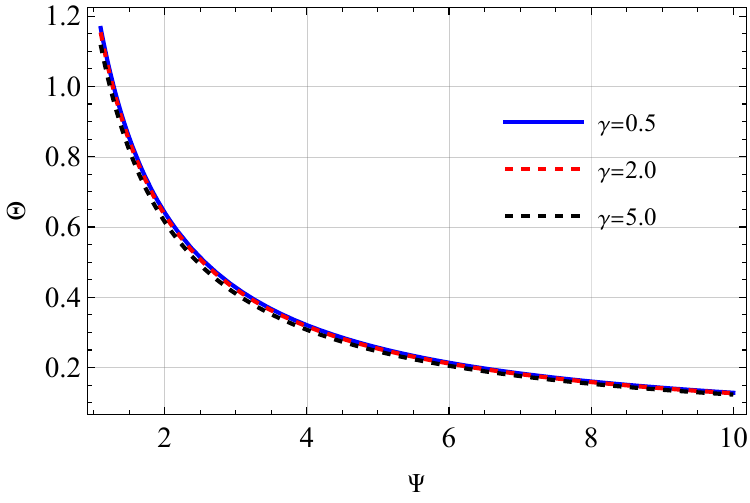}~~~~ \includegraphics[height=7cm,width=8cm]{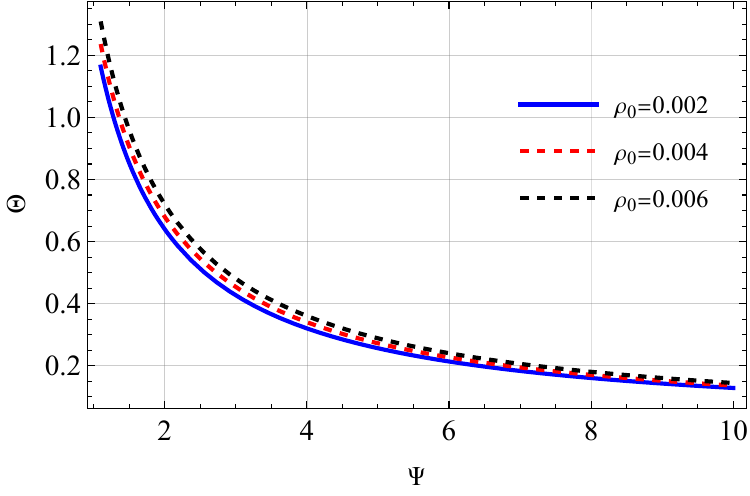}
    \caption{Variation of the deflection angle \(\Theta\) with \(\Psi\) for different coupling parameter \(\gamma\) (left) and density parameter \(\rho_0\) (right).}
    \label{fig05}
\end{figure*}
For a circular trajectory with constant radial coordinate \(R\), the radial component of the covariant derivative reduces to
\begin{equation}
\left(
\nabla_{\dot{\eta}_R}\dot{\eta}_R
\right)^r
=
\bar{\Gamma}^{r}_{\phi\phi}
\left(
\dot{\eta}_R^\phi
\right)^2+\dot{\eta}_R^\phi
\left(
\partial_\phi \dot{\eta}_R^r
\right)
.
\end{equation}

Here, the first term vanishes because $r$ is constant along the
circular curve $\eta_R$, while the second term can be evaluated using
the unit-speed condition
$\bar{g}_{\phi\phi}(\dot{\eta}_R^\phi)^2=1$. Consequently, in the
asymptotic limit $R\to\infty$, the geodesic curvature approaches
$\kappa(\eta_R)\to R^{-1}$. Furthermore, in the large-$R$ limit, the
jump angles at the observer ($\xi_O$) and the source ($\xi_S$)
both approach $\pi/2$, yielding
$\xi_O+\xi_S\to\pi$ \cite{Gibbons:2008rj}. Therefore, we obtain
the following relation
\begin{equation}
\lim_{R\rightarrow\infty}
\kappa(\eta_R)
\frac{dt}{d\phi}
\rightarrow 1.
\end{equation}

This result demonstrates that the optical metric approaches the
Euclidean form in the asymptotic limit. With the geodesic curvature
now established, the Gauss-Bonnet theorem can be expressed in the
following form
\begin{equation}
\begin{aligned}
\iint_{\mathcal{H}_R} \mathcal{K}\,dS
+
\int_{\eta_R} \kappa\,dt
&=
\lim_{R\rightarrow\infty}
\left(\int_{0}^{\pi+\hat{\theta}} d\phi+
\iint_{\mathcal{S}_{\infty}} \mathcal{K}\,dS
\right)
\\
&=\pi,
\end{aligned}
\end{equation}

from which the deflection angle $\hat{\theta}$ can be extracted as
\begin{equation}\label{e56}
\Theta
=
-\int_{0}^{\pi}
\int_{r=\frac{\Psi}{\sin\phi}}^{\infty}
\mathcal{K}\,dS.
\end{equation}

Here, \(\Psi\) denotes the impact parameter. The integration of Eq.~(\ref{e56}), with the derived curvature expression, leads to the following analytical result for the light-deflection angle
\begin{eqnarray}
  \Theta\approx  \frac{r_0}{2 \Psi } \left(\frac{3 \left(\gamma ^2-10 \pi \gamma +16 \pi ^2\right) r_0^2 \rho_0}{3 \pi -\gamma }+2\right).
\end{eqnarray}
 \begin{figure*}[!htp]
    \centering
\includegraphics[height=8.3cm,width=9cm]{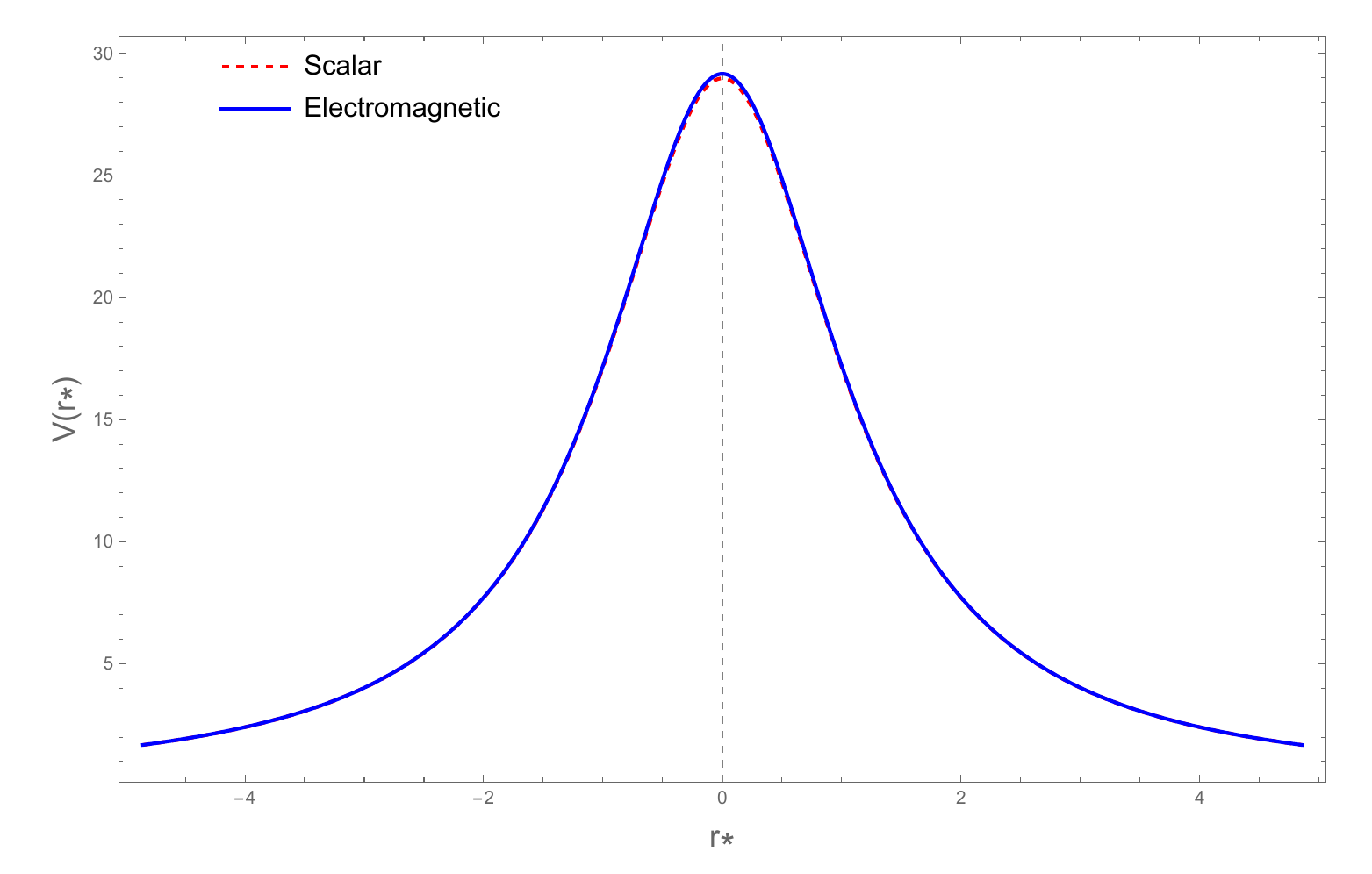}
   \caption{Scalar and electromagnetic perturbation potentials for the chosen values of the model parameters.}
    \label{fig06}
\end{figure*}
The optical behavior of the quartic wormhole solutions is investigated using the Gauss-Bonnet theorem, with particular emphasis on the dependence of the deflection angle \(\Theta\) on the impact parameter \(\Psi\). The effects of \(\gamma\) and \(\rho_0\) on the light-deflection profiles are presented in Fig.~\ref{fig05}. The results show that the gravitational bending of light gradually weakens as the impact parameter increases, irrespective of the parameter variation considered. The relatively large deflection at smaller impact parameters arises from the stronger influence of the wormhole spacetime on light passing through its immediate surroundings.\\

For the left panel of Fig.~\ref{fig05}, we consider \(\gamma=0.5\), \(2.0\), and \(5.0\) to study its effect on light deflection. The profiles show that larger values of \(\gamma\) lead to smaller deflection angles, whereas the strongest bending occurs for \(\gamma=0.5\). Although variations in \(\gamma\) produce different deflection angles, the differences become smaller at larger impact parameters. Thus, \(\gamma\) modifies the magnitude of the optical response without changing the general radial behaviour of light deflection. At larger values of \(\Psi\), the profiles approach each other, indicating a reduced influence of the coupling parameter on the optical behaviour of the outer region. A more noticeable variation is observed in the right panel of Fig.~\ref{fig05}, where larger values of \(\rho_0\) correspond to greater deflection angles. This trend is evident for the selected values \(0.002\), \(0.004\), and \(0.006\). A larger deflection angle is obtained for \(\rho_0=0.006\), while the \(\rho_0=0.002\) case exhibits the weakest bending. This behaviour holds across the entire plotted range, reflecting the influence of the density parameter on the optical response. The influence of the density parameter becomes more evident for smaller impact parameters, where changes in \(\rho_0\) produce larger differences in the deflection angle. As the impact parameter increases, the variation in the deflection angle due to \(\rho_0\) becomes less pronounced, indicating a weaker effect on light propagation at larger distances.\\

The comparison shows that increasing the coupling parameter \(\gamma\) weakens gravitational bending, whereas increasing the density parameter \(\rho_0\) strengthens it. The density-dependent curves show a greater separation than those associated with variations in \(\gamma\), indicating a stronger influence of \(\rho_0\) on the deflection angle over the considered ranges. Such behaviour confirms that the optical characteristics of the wormhole are affected by variations in the coupling and density parameters. The calculated deflection profiles reveal differences among the considered wormhole configurations. To establish whether these differences can be detected observationally, further analysis involving parameter sensitivity, observational uncertainties, and suitable reference geometries is necessary.

\section{Wormhole Perturbation}
\subsection{Scalar Perturbation}
We now consider a massless scalar field \(\epsilon\) propagating in the wormhole geometry, with its evolution described by the curved-spacetime Klein-Gordon equation
\begin{equation}
\Box \epsilon =
\frac{1}{\sqrt{-g}}\partial_i
\left(\sqrt{-g}\,g^{ij}\partial_j \epsilon\right)=0.
\end{equation}

For the solution of this equation, we employ the standard separation of variables technique, similar to that used in the hydrogen-atom problem. This allows the scalar field to be separated into angular and temporal-radial contributions
\begin{equation}\label{eq59}
\epsilon(t,r,\theta,\phi)
=\frac{1}{r}\sum_{p,q}\psi_p(t,r)\mathcal{Y}_{pq}(\theta,\phi),
\end{equation}
where $\mathcal{Y}_{pq}(\theta,\phi)$ denotes the spherical harmonics and
$\psi_p(t,r)$ captures the radial and temporal dependence of the scalar
field. Assuming a harmonic time dependence of the form
\begin{equation}
\psi_p(t,r)=\hat{\psi}_p(r)e^{-i\omega t}.
\end{equation}
With the scalar field separated into its angular and temporal-radial components, Eq.~(\ref{eq59}) takes the form of a one-dimensional Schrodinger-like equation:
\begin{equation}
\left(\frac{d^2}{d\mathcal{X}^2}+\omega^2-\mathcal{V}_p(r)\right)
\hat{\psi}_p(\mathcal{X})=0.
\end{equation}
For convenience, we define a new radial coordinate \(\mathcal{X}\), known as the tortoise coordinate, through
\begin{equation}
d\mathcal{X}=\pm\frac{e^{-\nu(r)/2}~dr}
{\sqrt{1-\dfrac{b(r)}{r}}}.
\end{equation}

The explicit expression for the effective potential \(\mathcal{V}_p(r)\) is
\begin{equation}
\mathcal{V}_p(r)=e^{\nu}\left[\frac{\nu'}{2r}\left(1-\frac{b}{r}\right)+
\frac{l(l+1)}{r^2}
-\frac{b'r-b}{2r^3}
\right].
\end{equation}
\subsection{Electromagnetic Perturbation}
In addition to scalar-field perturbations, we investigate electromagnetic perturbations in the wormhole spacetime using the curved-spacetime Maxwell equations.
\begin{equation}
\left(\sqrt{-g}g^{i\rho}g^{j\sigma}\mathcal{F}_{\rho\sigma}\right)_{,j}=0,
\label{eq:63}
\end{equation}
where $\mathcal{F}_{ij}=\partial_i \mathcal{C}_j-\partial_j \mathcal{C}_i$ denotes the
electromagnetic field strength tensor, and $\mathcal{C}_i$ is the four-potential.
Since the underlying spacetime metric in Eq.~(\ref{e18}) is static and spherically
symmetric, the vector potential $\mathcal{C}_i$ can be decomposed using vector
spherical harmonics as follows
\begin{equation}
\begin{aligned}
\mathcal{C}_i(t,r,\theta,\phi)
={}&\int d\omega\sum_{p,q}
\Bigg[
a_{pq}(r)e^{-i\omega t}
\begin{pmatrix}
0\\
0\\
\dfrac{1}{\sin\theta}\partial_\phi \mathcal{Y}_{pq}(\theta,\phi)\\
-\sin\theta\,\partial_\theta \mathcal{Y}_{pq}
\end{pmatrix}_{\mathrm{odd}}
\\[8pt]
&\quad +e^{-i\omega t}
\begin{pmatrix}
f_{pq}(r)\mathcal{Y}_{pq}\\
u_{pq}(r)\mathcal{Y}_{pq}\\
k_{pq}(r)\partial_\theta \mathcal{Y}_{pq}\\
k_{pq}(r)\partial_\phi \mathcal{Y}_{pq}
\end{pmatrix}_{\mathrm{even}}
\Bigg].
\end{aligned}
\label{eq:64}
\end{equation}

The decomposition separates the electromagnetic perturbations into two independent parity sectors: axial (odd) and polar (even). Their respective parity factors are \((-1)^{p+1}\) and \((-1)^p\), and their radial dependence is encoded in four functions, \(a_{pq}(r)\), \(f_{pq}(r)\), \(u_{pq}(r)\), and \(k_{pq}(r)\). Both sectors, however, share the same master equation for radial propagation
$\psi_{pq}^{(1)}$:
\begin{equation}
\frac{d^2\psi_{pq}^{(1)}}{d\mathcal{X}^2}
+\left(\omega^2-e^\nu\frac{p(p+1)}{r^2}\right)
\psi_{pq}^{(1)}=0,
\label{eq:65}
\end{equation}
with the corresponding effective potential
\begin{equation}
\mathcal{V}_l^{(1)}(r)=e^\nu\frac{p(p+1)}{r^2}.
\label{eq:66}
\end{equation}

Fig.~\ref{fig06} illustrates the effective potentials for scalar and
electromagnetic perturbations of the quartic wormhole. Both potentials
exhibit a positive, approximately symmetric barrier around the throat,
with a maximum of about $\mathcal{V}_{\max}\simeq 29$. Their similar profiles
indicate that both fields experience nearly identical scattering
environments, although the electromagnetic potential is slightly
higher near the peak. Consequently, electromagnetic waves may undergo
stronger reflection than scalar waves for frequencies near the barrier
maximum. The absence of a pronounced potential well suggests no
obvious region of stable wave trapping in the plotted configuration.

\begin{figure*}[!htp]
    \centering
\includegraphics[height=7.5cm,width=9.5cm]{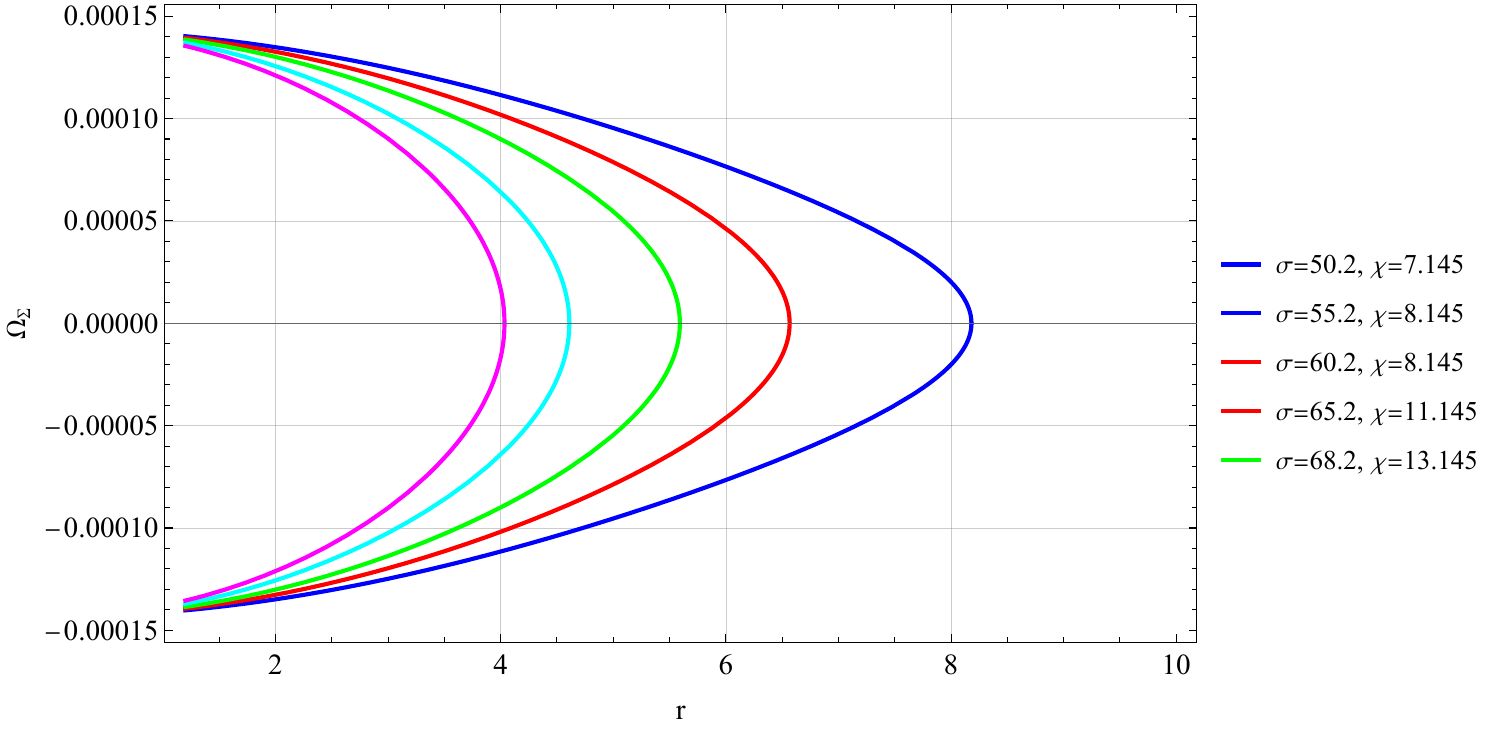}
    \caption{Variation of the redshift and blueshift with the radial coordinate for different values of the model parameters.}
    \label{fig07}
\end{figure*}
\section{Red Blue Shift Functions}
To explore the observational properties of the wormhole spacetime, we examine the redshift and blueshift of photons emitted by test particles orbiting near its throat.  The red-blue shift function provides a useful framework for examining the relationship between the emitter and the observer. For the wormhole geometry under consideration, this function can be determined using the following relation~\cite{Mustafa:2023mls}:
\begin{eqnarray} \label{e65}
\Omega_{\Sigma}(r)=\pm\sqrt{-\frac{g_{\phi\phi}}{g_{tt}}
\left(\frac{\mathcal{L}}{g_{\phi\phi}}-\frac{\sigma~\mathcal{E}_{\phi}}{g_{\phi\phi}}\right)},
\end{eqnarray} 
where $g_{tt}$ and $g_{\phi\phi}$ denote the temporal and azimuthal components of the spacetime metric, respectively, $\mathcal{L}$ represents the conserved angular momentum, $\sigma$ is the coupling parameter associated with the electromagnetic interaction, and $\mathcal{E}_{\phi}$ is the azimuthal component of the electromagnetic four-potential. To investigate the electromagnetic contribution to the frequency-shift behavior, we introduce the following phenomenological ansatz for the azimuthal potential:
\begin{eqnarray}
    \mathcal{E}_{\phi}(r,\theta)=
\frac{\mu r^2}{(r^2+\ell^2)^{3/2}}\sin^2\theta,
\end{eqnarray}

In the equatorial plane, \(\theta=\pi/2\), the azimuthal component of the electromagnetic four-potential reduces to
\begin{equation}
\mathcal{E}_{\phi}(r)=\frac{\mu r^2}{(r^2+\ell^2)^{3/2}}.
\end{equation}
In the asymptotic regime \(r\gg\ell\), the potential behaves as
\begin{equation}
\mathcal{E}_{\phi}(r)\sim\frac{\mu}{r},
\qquad r\rightarrow\infty.
\end{equation}
Here, \(\mu\) determines the amplitude of the potential, whereas \(\ell\) characterizes its radial scale. For simplicity, one may set \(\ell=r_0\), where \(r_0\) denotes the wormhole throat radius; however, this identification represents a modeling choice rather than a constraint imposed by the wormhole geometry.\\

Substituting the above potential into the red-blue shift function, Eq.~(\ref{e65}), and using \(g_{tt}=-e^{2\Phi_0}\) and \(g_{\phi\phi}=r^2\), we formally obtain
\begin{equation}
\Omega_{\Sigma}(r)=\pm e^{-\Phi_0}
\sqrt{\mathcal{L}-\frac{\sigma\mu r^2}{(r^2+\ell^2)^{3/2}}}.
\end{equation}

Fig.~\ref{fig07} shows how the angular velocity $\Omega_\Sigma(r)$ changes with
the radial distance $r$ for different values of the electromagnetic
parameters $\sigma$ and $\mu$. To study the electromagnetic effects on particle motion, we introduce an azimuthal potential in the background of the quartic-density wormhole geometry. The inclusion of the electromagnetic potential allows us to examine how the particle charge affects its motion in the wormhole background. For positive $\mu$ and $\ell$, the potential
starts from zero, increases with radial distance, reaches its maximum
at $r=\sqrt{2}\ell$, and then gradually decreases towards zero at large
distances. Therefore, the effect of the electromagnetic field is
strongest near this radial position and becomes weaker as the particle
moves farther away. The expression for $\Omega_\Sigma(r)$ also shows that the
electromagnetic contribution is combined with the angular-momentum
term through $\mathcal{L}-\sigma\mathcal{E}_{\phi}(r)$. This means that
the angular velocity depends on both the particle charge and the
strength of the electromagnetic potential. For positive $\sigma\mu$,
the electromagnetic term reduces the quantity inside the square root
and hence decreases the magnitude of the angular velocity, provided
the other parameters remain fixed and the angular velocity remains
real. The positive and negative branches in the figure represent
motion in opposite directions. These results show that the
electromagnetic field can change the circular motion of charged
particles in the wormhole spacetime. Since the motion of the emitting
particle also affects the frequency of the emitted light, the
variation of angular velocity is important for studying the redshift
and blueshift of radiation. Thus, the electromagnetic field provides
an additional way to examine the frequency-shift behaviour of the
considered quartic-density wormhole.

\section{Concluding Remarks}
By considering a quartic density profile in \(f(T,\tau)\) gravity, we obtain analytical solution describing traversable wormhole geometry. The derived shape function meets the essential geometrical requirements, ensuring a well-defined throat, the required flare-out behaviour, the absence of event horizons, and an asymptotically flat structure. The radial variation of the energy density and both pressure components reveals the anisotropic nature of the matter distribution. The energy-condition analysis indicates that the radial NEC is violated, whereas the tangential NEC is satisfied, highlighting the role of effective exotic matter in sustaining the wormhole geometry. The optical Gaussian curvature varies non-monotonically with \(r\), demonstrating how the parameter \(\gamma\) affects the geometrical properties of the optical spacetime. The deflection angle obtained through the GBT exhibits opposite trends with respect to the two parameters, decreasing with \(\gamma\) and increasing with \(\rho_0\). These findings provide a basis for examining the gravitational lensing behaviour of the wormhole under different choices of model parameters.\\

The scalar and electromagnetic effective potentials are also analysed to investigate wave propagation in the wormhole background. Both profiles display positive and approximately symmetric barriers near the throat, attaining maximum values of about \(\mathcal{V}_{\max}\simeq 29\). Their similar profiles indicate comparable wave-scattering behaviour, although the slightly higher electromagnetic barrier suggests stronger reflection for suitable frequencies. The absence of a pronounced potential well indicates no evident region of stable wave trapping in the examined configuration; however, a complete assessment of dynamical stability requires further investigation of the perturbation equations and their mode spectrum. The angular velocity $\Omega_\Sigma(r)$ was examined for different
values of the electromagnetic parameters $\sigma$ and $\mu$. The
results show that the electromagnetic potential affects the circular
motion of charged particles through the term
$\mathcal{L}-\sigma\mathcal{E}_{\phi}(r)$. Its influence is strongest
near $r=\sqrt{2}\ell$ and gradually decreases at larger radial
distances. The positive and negative branches of the angular velocity
represent opposite directions of motion. These changes in angular
velocity also play an important role in the redshift and blueshift
analysis, as the frequency shift depends on the motion of the emitting
particle. Overall, the present study establishes a unified framework for examining the geometrical structure, matter distribution, energy conditions, gravitational lensing, and wave-propagation properties of quartic wormholes in $f(T,\tau)$ gravity. These findings motivate further studies of their physical viability, dynamical stability, and potential observational signatures.

\section*{Acknowledgments}

S. K. Maurya gratefully acknowledges the unwavering support and encouragement of the administration of the University of Nizwa, Sultanate of Oman. Sourav Chaudhary sincerely acknowledges the support extended by Lovely Professional University. Jitendra Kumar expresses his sincere gratitude to the Department of Mathematics, Central University of Haryana, for its continued support.
 
\section*{Data Availability Statement}
This manuscript has no associated data or the data will not be deposited (There is no observational data related to this article. The necessary calculations and graphic discussion are already available in the manuscript.)

\section*{Conflict of Interest}
The authors declare no conflict of interest.


\begin{thebibliography}{99}
\bibitem{Brown:2019hmk}
A.~R.~Brown, H.~Gharibyan, S.~Leichenauer, H.~W.~Lin, S.~Nezami, G.~Salton, L.~Susskind, B.~Swingle and M.~Walter,
PRX Quantum \textbf{4}, no.1, 010320 (2023)
\bibitem{Nezami:2021yaq}
S.~Nezami, H.~W.~Lin, A.~R.~Brown, H.~Gharibyan, S.~Leichenauer, G.~Salton, L.~Susskind, B.~Swingle and M.~Walter,
PRX Quantum \textbf{4}, no.1, 010321 (2023)
\bibitem{Xavier:2024iwr}
S.~V.~M.~C.~B.~Xavier, C.~A.~R.~Herdeiro and L.~C.~B.~Crispino,
Phys. Rev. D \textbf{109}, no.12, 124065 (2024)
\bibitem{Mustafa:2023kqt}
G.~Mustafa, S.~K.~Maurya and S.~Ray,
Fortsch. Phys. \textbf{71}, no.6-7, 2200129 (2023)
\bibitem{Mustafa:2022fxn}
G.~Mustafa, S.~K.~Maurya and S.~Ray,
Astrophys. J. \textbf{941}, no.2, 170 (2022)
\bibitem{Mustafa:2023qxf}
G.~Mustafa, S.~K.~Maurya, S.~Ray and F.~Javed,
Annals Phys. \textbf{460}, 169551 (2024)
\bibitem{Jahnke:2019gxr}
V.~Jahnke, K.~Y.~Kim and J.~Yoon,
JHEP \textbf{05}, 037 (2019)
\bibitem{Maldacena:2020sxe}
J.~Maldacena and A.~Milekhin,
Phys. Rev. D \textbf{103}, no.6, 066007 (2021)
\bibitem{Molla:2024loc}
N.~U.~Molla, H.~Chaudhary, U.~Debnath, G.~Mustafa and S.~K.~Maurya,
Eur. Phys. J. C \textbf{85}, no.1, 15 (2025)
\bibitem{Bronnikov:2002rn}
K.~A.~Bronnikov and S.~W.~Kim,
Phys. Rev. D \textbf{67}, 064027 (2003)
\bibitem{Wang:2017cnd}
D.~Wang and X.~H.~Meng,
Front. Phys. (Beijing) \textbf{13}, no.1, 139801 (2018)
\bibitem{Aounallah:2020rlf}
H.~Aounallah, A.~R.~Soares and R.~L.~L.~Vit{\'o}ria,
Eur. Phys. J. C \textbf{80}, no.5, 447 (2020)
\bibitem{Zhang:2024sgs}
J.~Zhang and Y.~Xie,
Phys. Rev. D \textbf{109}, no.4, 043032 (2024)
\bibitem{Huang:2023yqd}
H.~Huang, J.~Kunz, J.~Yang and C.~Zhang,
Phys. Rev. D \textbf{107}, no.10, 104060 (2023)
\bibitem{Muniz:2026qmb}
C.~R.~Muniz, J.~A.~Rebou{\c{c}}as, M.~B.~Cruz and R.~M.~P.~Neves,
[arXiv:2607.21720 [gr-qc]].
\bibitem{Jaramillo:2023pny}
V.~Jaramillo, M.~Lira, D.~Mart{\'\i}nez-Carbajal and D.~N{\'u}{\~n}ez,
Phys. Rev. D \textbf{109}, no.6, 064007 (2024)
\bibitem{Nojiri:2023dvf}
S.~Nojiri and G.~G.~L.~Nashed,
Phys. Rev. D \textbf{108}, no.12, 124049 (2023)
\bibitem{Capozziello:2011hj}
S.~Capozziello, V.~F.~Cardone, H.~Farajollahi and A.~Ravanpak,
Phys. Rev. D \textbf{84}, 043527 (2011)
\bibitem{Krssak:2018ywd}
M.~Krssak, R.~J.~van den Hoogen, J.~G.~Pereira, C.~G.~B{\"o}hmer and A.~A.~Coley,
Class. Quant. Grav. \textbf{36}, no.18, 183001 (2019)
\bibitem{Boehmer:2012uyw}
C.~G.~Boehmer, T.~Harko and F.~S.~N.~Lobo,
Phys. Rev. D \textbf{85}, 044033 (2012)
\bibitem{Jamil:2011ptc}
M.~Jamil, D.~Momeni, M.~Raza and R.~Myrzakulov,
Eur. Phys. J. C \textbf{72}, 1999 (2012)
\bibitem{Harko:2014aja}
T.~Harko, F.~S.~N.~Lobo, G.~Otalora and E.~N.~Saridakis,
JCAP \textbf{12}, 021 (2014)
\bibitem{Wu:2010mn}
P.~Wu and H.~W.~Yu,
Phys. Lett. B \textbf{693}, 415-420 (2010)
\bibitem{Nunes:2018evm}
R.~C.~Nunes, S.~Pan and E.~N.~Saridakis,
Phys. Rev. D \textbf{98}, no.10, 104055 (2018)
\bibitem{Geng:2011aj}
C.~Q.~Geng, C.~C.~Lee, E.~N.~Saridakis and Y.~P.~Wu,
Phys. Lett. B \textbf{704}, 384-387 (2011)
\bibitem{Wei:2011yr}
H.~Wei,
Phys. Lett. B \textbf{712}, 430-436 (2012)
\bibitem{Bamba:2013ooa}
K.~Bamba, S.~Capozziello, M.~De Laurentis, S.~Nojiri and D.~S{\'a}ez-G{\'o}mez,
Phys. Lett. B \textbf{727}, 194-198 (2013)
\bibitem{Arora:2022jso}
S.~Arora, A.~Bhat and P.~K.~Sahoo,
Fortsch. Phys. \textbf{71}, no.2-3, 2200162 (2023)
\bibitem{Reboucas:2026fxc}
J.~A.~Rebou{\c{c}}as and E.~Otoniel,
[arXiv:2609.12000 [gr-qc]].
\bibitem{Ditta:2021uoe}
A.~Ditta, I.~Hussain, G.~Mustafa, A.~Errehymy and M.~Daoud,
Eur. Phys. J. C \textbf{81}, no.10, 880 (2021)
\bibitem{Muniz:2026akg}
C.~R.~Muniz, R.~Avalos, J.~A.~Rebou{\c{c}}as, F.~B.~Lustosa and F.~T.~B.~Sampaio,
[arXiv:2609.17597 [gr-qc]].

\bibitem{Reboucas:2026fwz}
J.~A.~Rebou{\c{c}}as, E.~Otoniel, M.~C.~Ara{\'u}jo, J.~Furtado and C.~R.~Muniz,
[arXiv:2607.26882 [gr-qc]].
\bibitem{Mustafa:2021vqz}
G.~Mustafa, M.~Ahmad, A.~{\"O}vg{\"u}n, M.~F.~Shamir and I.~Hussain,
Fortsch. Phys. \textbf{69}, no.8-9, 2100048 (2021)


\bibitem{Harko:2011kv}
T.~Harko, F.~S.~N.~Lobo, S.~Nojiri and S.~D.~Odintsov,
Phys. Rev. D \textbf{84}, 024020 (2011)
doi:10.1103/PhysRevD.84.024020
\bibitem{William}
Hoover, William G., Steven G. Gray, and Keith W. Johnson. 
The Journal of Chemical Physics 55.3 (1971): 1128-1136.
\bibitem{Gibbons:2008rj}
G.~W.~Gibbons and M.~C.~Werner,
Class. Quant. Grav. \textbf{25}, 235009 (2008)
doi:10.1088/0264-9381/25/23/235009
\bibitem{Werner:2012rc}
M.~C.~Werner,
Gen. Rel. Grav. \textbf{44}, 3047-3057 (2012)
doi:10.1007/s10714-012-1458-9
\bibitem{Mustafa:2023mls}
G.~Mustafa,
Phys. Lett. B \textbf{848}, 138407 (2024)
doi:10.1016/j.physletb.2023.138407


















 \end{thebibliography}
\end{document}